\documentclass[english]{revtex4}
\usepackage[T1]{fontenc}
\usepackage[latin1]{inputenc}
\usepackage{graphicx}
\usepackage{amssymb}

\makeatletter

\usepackage{babel}
\makeatother
\begin{document}

\title{Momentum Dynamics of One Dimensional Quantum Walks }

\author{Ian Fuss}

\email{Ian.Fuss@dsto.defence.gov.au}

\affiliation{Defence Science and Technology Organisation (DSTO), Edinburgh, Australia.}

\affiliation{School of Electrical and Electronic Engineering, University of Adelaide,
Australia.}

\author{Langford B. White and Sanjeev Naguleswaran}

\affiliation{School of Electrical and Electronic Engineering, University of Adelaide,
Australia.}

\author{Peter J. Sherman}

\affiliation{Department of Aerospcae Engineering, Iowa State University, Ames,
Iowa.}

\begin{abstract}
We derive the momentum space dynamic equations and state functions
for one dimensional quantum walks by using linear systems and Lie
group theory. The momentum space provides an analytic capability similar
to that contributed by the z transform in discrete systems theory.
The state functions at each time step are expressed as a simple sum
of three Chebyshev polynomials. The functions provide an analytic
expression for the development of the walks with time.
\end{abstract}
\maketitle

\section{Introduction}

The study of quantum walks has received considerable attention since
the introductory papers on the subject, such as \cite{aharanov00,Kempe03}
and references therein. In this paper, we develop an analytic approach
to study the properties of these walks based on a momentum space representation.

This paper is structured such that in Section 2 of the paper the momentum
space dynamic equations for one dimensional quantum walks are derived
via the Z transform of the position space dynamic equations and its
representation of the discrete Fourier transform when Z lies on the
unit circle. An exponential form of of the momentum space time operator
is derived in section 3 by using the group theory of $SU(2)$ and
a matrix inner product space. The exponential form allows a simple
analytic calculation of the time evolution operator for arbitrary
time intervals. This is used in Section 4 to obtain analytic expressions
for the momentum space wave functions of quantum walks at arbitrary
times. These wave functions are expressed quite simply in terms of
Chebyshev Polynomials of the second kind. Some plots of the momentum
space probability densities for different parameter values and times
are provided in section 5. The conclusions are summarised in Section
6.

\section{Momentum Space Dynamic Equations}

For a given $\psi(0,0)$ we consider the evolution of a quantum state
$\psi(t,x)\in C^{2}$ for discrete times $t\ge0$ on a line $x\in Z.$
The dynamics of the state then evolve according to the difference
equations,\begin{eqnarray}
 & \psi_{0}(t,x)=e^{i\alpha}[a\psi_{0}(t-1,x-1)+b\psi_{1}(t-1,x-1)],\nonumber \\
 & \psi_{1}(t,x)=e^{i\alpha}[-b^{*}\psi_{0}(t-1,x+1)+a^{*}\psi_{1}(t-1,x+1)],\label{eq:momdy}\end{eqnarray}
where $|a|^{2}+|b|^{2}=1$ and $\alpha\in R$.

Taking two-dimensional $Z$ transforms of these equations yields\begin{eqnarray}
 & \psi_{0}(z_{1},z_{2})=e^{i\alpha}z_{1}^{-1}z_{2}^{-1}[a\psi_{0}(z_{1},z_{2})+b\psi_{1}(z_{1},z_{2})\nonumber \\
 & \psi_{1}(z_{1},z_{2})=e^{i\alpha}z_{1}^{-1}z_{2}^{-1}[-b^{*}\psi_{0}(z_{1},z_{2})+a^{*}\psi_{1}(z_{1},z_{2}).\end{eqnarray}

Thus the transfer matrix for the system is\begin{equation}
B(z_{1},z_{2})=e^{i\alpha}z_{1}^{-1}\left[\begin{array}{cc}
az_{2}^{-1} & bz_{2}^{-1}\\
-b^{\ast}z_{2} & a^{\ast}z_{2}\end{array}\right]\end{equation}

therefore, for any iteration (time) index $n$, the quantum walk state
$\ \Psi(n,x)$ has transform $x\leftrightarrow z$ \begin{equation}
\Psi(n,x)\leftrightarrow e^{in\alpha}C^{n}(z)\Psi(0,0),\end{equation}

where $C(z)$ is the matrix polynomial\begin{equation}
C(z)=\left[\begin{array}{cc}
az^{-1} & bz^{-1}\\
-b*z & a*z\end{array}\right].\end{equation}
 It should be noted that $C$ is paraunitary, that is $C^{-1}(z)=C^{T}(1/z).$
In particular this implies that $C(z)$ is unitary on $|z|=1.$ Further
we note that $detC(e^{ip)})=1$ and hence the matrix \begin{equation}
S(p)=C(e^{ip})\label{eq:6}\end{equation}

is unimodular. The Fourier transform $x\leftrightarrow p$ is\begin{equation}
\Psi(n,x)\leftrightarrow e^{in\alpha}S^{n}(p)\Psi(0,0).\end{equation}

Thus by choosing Planck's constant $\hbar=1,$ the momentum space
representation of the quantum walk state vector $\phi(n,p)$ evolves
as\begin{equation}
\phi(n,p)=e^{in\alpha}S^{n}(p)\phi(0,p),\label{eq:8}\end{equation}

where\begin{equation}
\phi(0,p)=\psi(0,0)=\left[\begin{array}{c}
\psi_{0}(0,0)\\
\psi_{1}(0,0)\end{array}\right].\label{eq:9}\end{equation}

Thus the time evolution operator in the momentum space is a $2\times2$
matrix polynomial. Hence, the momentum space equations are much more
amenable to analysis than those in position space.

\section{Exponentiation of the Time Evolution Operator}

The unimodular matrix $S(p)$ can be written in exponential form as\begin{equation}
S(p)=Exp(i\theta(p)\overrightarrow{c}(p).\overrightarrow{\sigma})\label{eq10}\end{equation}

where $\theta$ and $\overrightarrow{c}$ are real functions of $p$
and the matrix vector $\overrightarrow{\sigma}$ has Pauli matrix
components \cite{merz}\[
\sigma_{1}=\left[\begin{array}{cc}
0 & 1\\
1 & 0\end{array}\right],\]
\[
\sigma_{2}=\left[\begin{array}{cc}
0 & -i\\
i & 0\end{array}\right]\]
and\begin{equation}
\sigma_{3}=\left[\begin{array}{cc}
1 & 0\\
0 & -1\end{array}\right].\label{eq:11}\end{equation}

The inner product\[
(A,B)=\frac{1}{2}Tr(AB)\]

defined on the vector space of $2\times2$ unitary matrices gives
an inner product space. The set of matrices $\{ I,\sigma_{1},\sigma_{2,}\sigma_{3}\},$
provide an ortho-normal basis for this space. 

The coefficients of the matrices can be evaluated by taking the inner
product of both sides of (\ref{eq10}) \[
(\sigma_{i},S(p))=(\sigma_{i},Exp(i\theta(p)\overrightarrow{c}(p).\overrightarrow{\sigma})\]
with each of the matrices $\sigma_{i}.$ In doing this we note that
a generalised de-Moivre principle gives\[
Exp(i\theta\overrightarrow{c}.\overrightarrow{\sigma})=Icos(\theta)+i\overrightarrow{c}.\overrightarrow{\sigma}sin(\theta),\]
where the $p$ dpendence has been suppressed for simplicity. Hence,\begin{equation}
(I,Exp(i\theta\overrightarrow{c}.\overrightarrow{\sigma}))=cos(\theta)\label{eq13}\end{equation}
and\begin{equation}
(\sigma_{j},Exp(i\theta\overrightarrow{c}.\overrightarrow{\sigma}))=ic_{j}sin(\theta).\label{eq14}\end{equation}

The equivalent coefficients for $S(p)$ can be obtained by defining
\[
a=cos(\beta)e^{-i\gamma},\]
\begin{equation}
b=sin(\beta)e^{-i\delta}.\label{eq:define_ab}\end{equation}

Substituting in (\ref{eq:6}) gives\begin{equation}
S(p)=\left[\begin{array}{cc}
cos(\beta)e^{-i(p+\gamma)} & sin(\beta)e^{-i(p+\delta)}\\
-sin(\beta)e^{i(p+\delta)} & cos(\beta)e^{i(p+\gamma)}\end{array}\right].\end{equation}

These expressions can be simplified by setting $p'=p+\gamma$ and
$p''=p+\delta$. Using de Moivre's principle once again we obtain
the transition matrix coefficients\begin{eqnarray}
 & (I,S(p))=cos(\beta)cos(p'),\nonumber \\
 & (\sigma_{1},S(p))=-isin(\beta)sin(p''),\nonumber \\
 & (\sigma_{2},S(p))=isin(\beta)cos(p''),\nonumber \\
 & (\sigma_{3},S(p))=-icos(\beta)sin(p').\label{eq17}\end{eqnarray}

Comparing coefficients in equations (\ref{eq13}) and (\ref{eq14})
with those of (\ref{eq17}) we obtain\[
cos(\theta)=cos(\beta)cos(p'),\]
\[
c_{1}sin(\theta)=-sin(\beta)sin(p''),\]
\[
c_{2}sin(\theta)=sin(\beta)cos(p''),\]
\begin{equation}
c_{3}sin(\theta)=-cos(\beta)sin(p').\label{eq20}\end{equation}

\section{Momentum Space State Functions}

A dynamic equation for momentum space state functions was given in
(\ref{eq:8}). The exponentiation of the operator in (\ref{eq10})
enables us to write the powers of the evolution operator as\begin{equation}
S^{n}(p)=Exp(in\theta\overrightarrow{c}.\overrightarrow{\sigma})=Icos(n\theta)+i\overrightarrow{c}.\overrightarrow{\sigma}sin(n\theta).\end{equation}

The trigonometric expressions in the above equation can be expressed
in terms of the Chebyshev polynomials $T_{n}$ and $U_{n}$ as \cite{Arf}\[
cos(n\theta)=T_{n}(cos(\theta))\]
and\begin{equation}
sin(n\theta)=U_{n-1}(cos(\theta))sin(\theta).\end{equation}

Using these expressions and writing the dot product as a sum of components
(\ref{eq:11}) becomes\begin{equation}
S^{n}(p)=T_{n}(cos(\theta))I+iU_{n-1}(cos(\theta))\sum_{i=1}^{3}c_{i}sin(\theta)\sigma_{i}.\end{equation}

The equalities of (\ref{eq20}) enable us to rewrite this as\begin{equation}
S^{n}(p)=T_{n}(cos(\beta)cos(p'))I-iU_{n-1}(cos(\beta)cos(p'))[sin(\beta)sin(p'')\sigma_{1}-sin(\beta)cos(p'')\sigma_{2}+cos(\beta)sin(p')\sigma_{3}]\end{equation}
Using the Pauli matrices the matrix polynomial\begin{eqnarray}
 & S^{n}(p)=\left[\begin{array}{cc}
T_{n}(cos(\beta)cos(p') & U_{n-1}(cos(\beta)cos(p'))sin(\beta)cos(p'')\\
-U_{n-1}(cos(\beta)cos(p'))sin(\beta)cos(p'') & T_{n}(cos(\beta)cos(p'))\end{array}\right]\nonumber \\
- & i\left[\begin{array}{cc}
U_{n-1}(cos(\beta)cos(p'))cos(\beta)sin(p') & U_{n-1}(cos(\beta)cos(p'))sin(\beta)sin(p'')\\
U_{n-1}(cos(\beta)cos(p'))sin(\beta)sin(p'') & -U_{n-1}(cos(\beta)cos(p'))cos(\beta)sin(p')\end{array}\right]\label{eq25}\end{eqnarray}
 is obtained.

The evolution of the quantum walk in momentum space representation
given in (\ref{eq:8} )can also be expressed as \begin{equation}
\phi(n,p)e^{-in\alpha}=S^{n}(p)\phi(0,p).\label{eq:21}\end{equation}

(\ref{eq25}) and (\ref{eq:9}) enable this expression to be written
as

\begin{eqnarray}
 & \phi_{0}(n,p)e^{-in\alpha}=[T_{n}(cos(\beta)cos(p'))-iU_{n-1}(cos(\beta)cos(p'))cos(\beta)sin(p')]\Psi_{0}(0,0)\nonumber \\
 & +[U_{n-1}(cos(\beta)cos(p'))sin(\beta)cos(p'')-iU_{n-1}(cos(\beta)cos(p'))sin(\beta)sin(p'')]\Psi_{1}(0,0)\end{eqnarray}

\begin{eqnarray}
 & \phi_{1}(n,p)e^{-in\alpha}=-[U_{n-1}(cos(\beta)cos(p'))sin(\beta)cos(p'')+iU_{n-1}(cos(\beta)cos(p'))sin(\beta)sin(p'')]\Psi_{0}(0,0)\nonumber \\
 & +T_{n}(cos(\beta)cos(p'))+iU_{n-1}(cos(\beta)cos(p'))cos(\beta)sin(p')]\Psi_{1}(0,0)\end{eqnarray}

By using the relation\begin{equation}
T_{n}(x)=U_{n}(x)-xU_{n-1}(x)\end{equation}

this can be written as

\begin{eqnarray}
 & \phi_{0}(n,p)e^{-in\alpha}=[U_{n}(cos(\beta)cos(p'))-U_{n-1}(cos(\beta)cos(p'))cos(\beta)[cos(p')+isin(p')]]\Psi_{0}(0,0)\nonumber \\
 & +[[U_{n-1}(cos(\beta)cos(p'))sin(\beta)[cos(p'')-sin(p'')]]\Psi_{1}(0,0)\end{eqnarray}
\begin{eqnarray}
 & \phi_{1}(n,p)e^{-in\alpha}=-[U_{n-1}(cos(\beta)cos(p'))sin(\beta)[cos(p'')+isin(p'')]]\Psi_{0}(0,0)\nonumber \\
 & +[U_{n}(cos(\beta)cos(p'))-U_{n-1}(cos(\beta)cos(p'))cos(\beta)[cos(p')+isin(p')]]\Psi_{1}(0,0).\end{eqnarray}

Inverting the de Moivre formula and moving the global phase term to
the right hand side gives the analytic expressions\begin{eqnarray}
 & \phi_{0}(n,p)=e^{in\alpha}[U_{n}(cos(\beta)cos(p'))-U_{n-1}(cos(\beta)cos(p'))cos(\beta)e^{ip}]\Psi_{0}(0,0)\nonumber \\
 & +e^{in\alpha}[U_{n-1}(cos(\beta)cos(p'))sin(\beta)e^{-ip}]\Psi_{1}(0,0)\end{eqnarray}
\begin{eqnarray}
 & \phi_{1}(n,p)=-e^{in\alpha}[U_{n-1}(cos(\beta)cos(p'))e^{ip}]\Psi_{0}(0,0)\nonumber \\
 & +e^{in\alpha}[U_{n}(cos(\beta)cos(p'))-U_{n-1}(cos(\beta)cos(p'))cos(\beta)e^{-ip}]\Psi_{1}(0,0)\end{eqnarray}

for the general momentum space state functions for a one dimensional
quantum walk at time n.

\section{Momentum Space Densities}

\begin{figure}
\begin{centering}\includegraphics[scale=0.6]{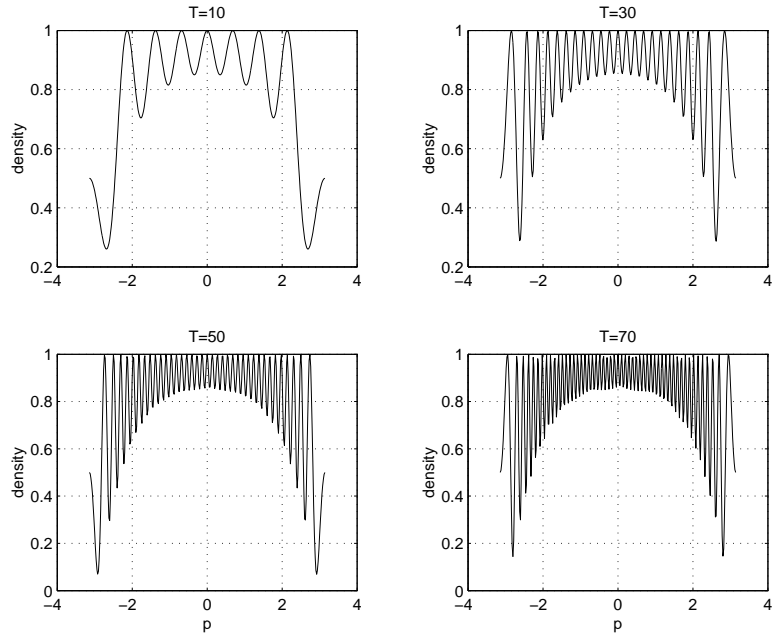}\par\end{centering}

\caption{\label{cap:Momentum-Space-Density1}Momentum Space Density functions
for $\beta=\frac{\pi}{8}$}
\end{figure}

\begin{figure}
\begin{centering}\includegraphics[scale=0.6]{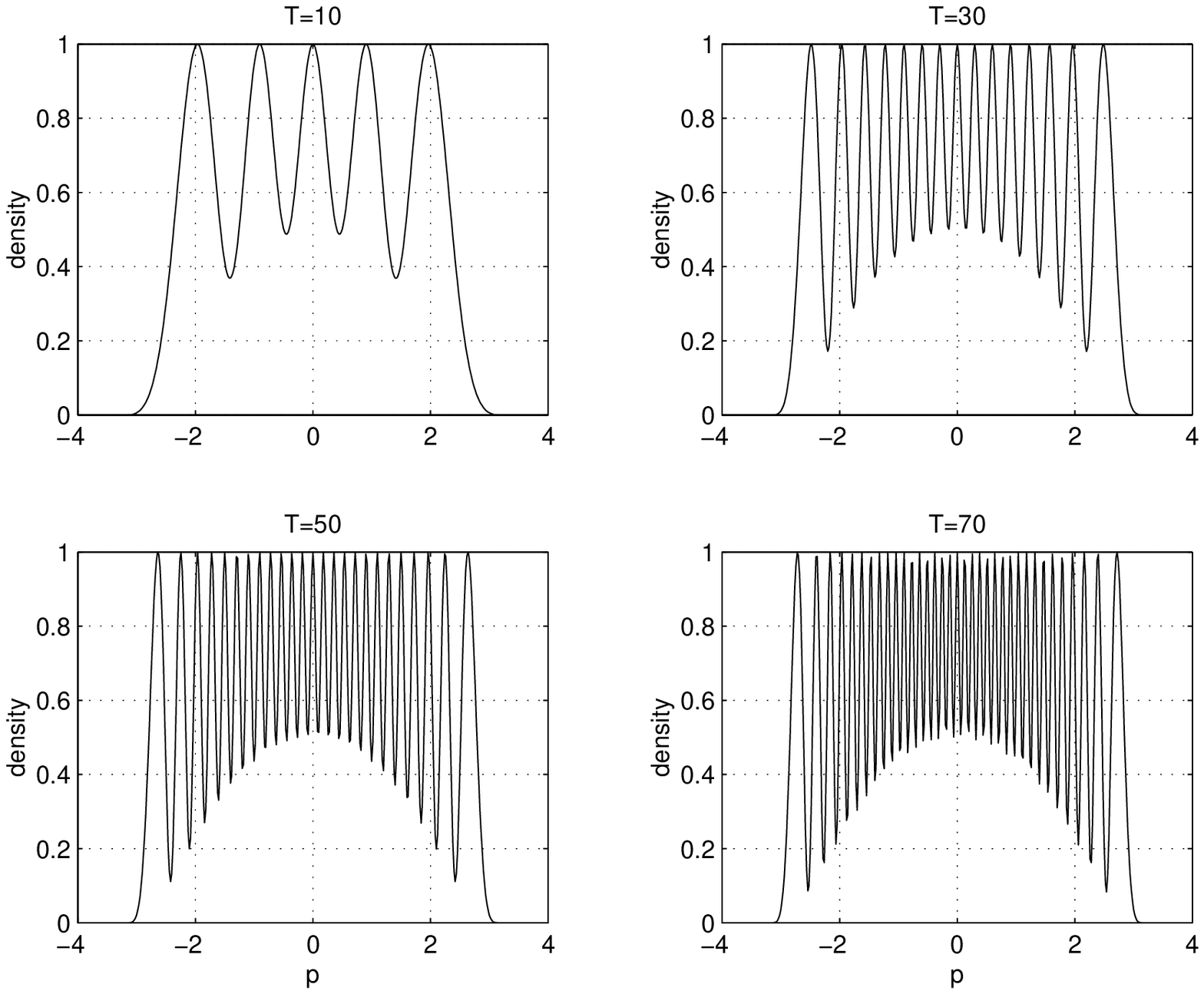}\par\end{centering}

\caption{\label{cap:Momentum-Space-Density2}Momentum Space Density functions
for $\beta=\frac{\pi}{4}$}
\end{figure}

\begin{figure}
\begin{centering}\includegraphics[scale=0.6]{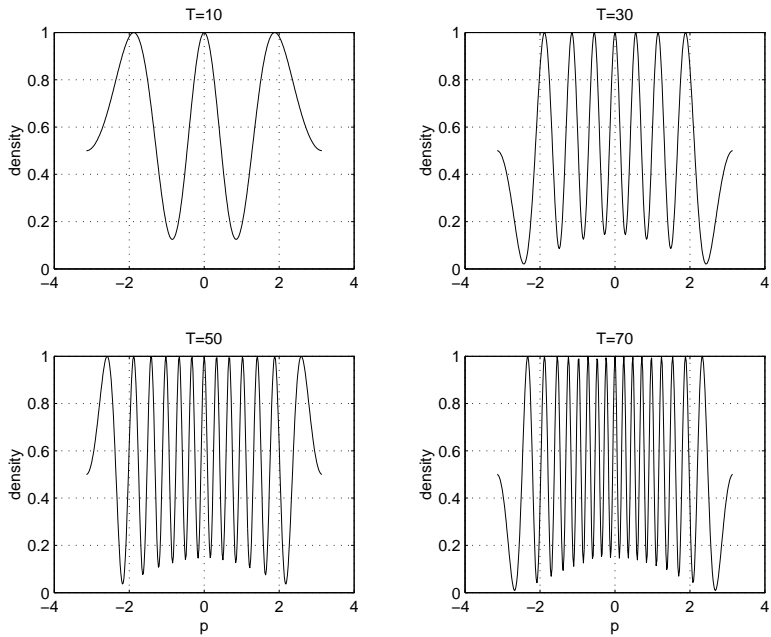}\par\end{centering}

\caption{\label{cap:Momentum-Space-Density3}Momentum Space Density functions
for $\beta=\frac{3\pi}{8}$}
\end{figure}

The denisity $|\phi_{0}(p:t)|^{2}$ for $\alpha=\gamma=\delta=0,$
$\Psi_{0}(0,0)=1$ and $\Psi_{1}(0,0)=0$ is plotted in figures \ref{cap:Momentum-Space-Density1},
\ref{cap:Momentum-Space-Density2}, \ref{cap:Momentum-Space-Density3}
for $\beta=\frac{\pi}{8},\frac{\pi}{4}$ and $\frac{3\pi}{8}$ and
for times $t=10,30,50,70.$ When $\beta$ is fixed the dominant feature
of the time series is an increase in oscillation frequency with time.
This corresponds to the increase in support of the position space
densities with time. The effect of increasing $\beta$ is to trade
a decrease in the constant component of the density function for an
increase in the oscillatory component. This corresponds to a shift
in the position space of probability density from the zero region
of the walk to the outer edges of the walk.

The sequences shows that the densities converge to a limit as time
increases. They also illustrate the fact that the momentum space is
an attractive representation in which to derive this limit because
the domain of the wave functions is constant, $p\in[-\pi,\pi].$ This
is in contrast to the real space where the domain expands with time.

\section{Conclusions}

It has been shown that the momentum space dynamic equations for a
quantum walk can be derived using a z transform of the position space
equations for the dynamic walk. An exponential representation of the
momentum space time evolution operator was derived by using Lie group
theory. This enabled the calculation of general momentum space wave
functions in terms of Chebyshev polynomials. Some simple calculations
of the momentum space probability densities illustrate the convergence
of the momentum wave functions to a limit as time increases. 

\bibliographystyle{plunsrt}
\bibliography{D:/DPOLP/documents/bib/quant}

\end{document}